\documentclass[aps,prl,twocolumn,superscriptaddress]{revtex4}

\usepackage{graphicx}
\usepackage[german,english]{babel}

\newcommand{\imp}{\ensuremath{\text{imp}}}

\newcommand{\ang}{\ensuremath{\text{\AA}}}

\begin{document}

\title{Resonant scattering by realistic impurities in graphene}

\author{T. O. Wehling}
\email{twehling@physnet.uni-hamburg.de}
\affiliation{I. Institut f{\"u}r Theoretische Physik, Universit{\"a}t Hamburg, Jungiusstra{\ss}e 9, D-20355 Hamburg, Germany}
\author{S. Yuan}
\affiliation{Radboud University of Nijmegen, Institute for
Molecules and Materials, Heijendaalseweg 135, 6525 AJ Nijmegen,
The Netherlands}
\author{A. I. Lichtenstein}
\affiliation{I. Institut f{\"u}r Theoretische Physik,
Universit{\"a}t Hamburg, Jungiusstra{\ss}e 9, D-20355 Hamburg,
Germany}
\author{A. K. Geim}
\affiliation{School of Physics and Astronomy, University of Manchester, Manchester M13 9PL, United Kingdom}
\author{M. I. Katsnelson}
\affiliation{Radboud University of Nijmegen, Institute for
Molecules and Materials, Heijendaalseweg 135, 6525 AJ Nijmegen,
The Netherlands}



\date{\today}

\begin{abstract}
We develop a first-principles theory of resonant impurities
in graphene and show that a broad range of typical realistic
impurities leads to the characteristic sublinear dependence of the
conductivity on the carrier concentration. By means of
density functional calculations various organic groups as well as
ad-atoms like H absorbed to graphene are shown to create midgap
states within $\pm 0.03$\,eV around the neutrality point. A low
energy tight-binding (TB) description is mapped out. Boltzmann
transport theory as well as a numerically exact Kubo formula
approach yield the conductivity of graphene contaminated with
these \textit{realistic} impurities in accordance with recent
experiments.
\end{abstract}
\maketitle

The mechanism determining the charge carrier mobility of present
graphene samples is being controversially debated. The main
experimental fact requiring an explanation is that, away from the
neutrality point, the conductivity of graphene is weakly
temperature dependent and approximately proportional to the
carrier concentration $n_e$ \cite{kostya2,kim}. This definitely
requires the assumption of some long-range interactions with scattering centers. The Coulomb interaction with charge impurities
is an ``explanation by default''
\cite{CoulombScatteringPapers}. However, it seems that some
experimental data cannot be explained in this way, especially, a
relatively weak sensitivity of the electron mobility to dielectric
screening \cite{Mohiuddin09}. Thus, alternative scattering
mechanisms are also discussed, such as frozen ripples \cite{KG08}
and resonant scatterers \cite{OGM06,KG08,KN07,SPG07}. In the first case
the long-range character of the interactions is due to the long-range
character of elastic deformations and in the second one due to
divergence of the scattering length. New experimental data
\cite{Nietal10} seem to support the latter possibility.

Theoretically, both suggestions face with serious problems. The
``ripple'' mechanism requires quenching of
the thermal bending fluctuations \cite{KG08,Ripples_EPL08}, but there are still
no realistic scenarios of such a quenching. Resonant
scattering naturally appears for vacancies \cite{SPG07} but they
do not exist, in noticeable concentrations, in graphene samples if
they are not created artificially, e.g., by irradiation
\cite{Chetal09}. Adsorbates on graphene can provide resonances
(quasilocalized states) close enough to the neutrality point
\cite{Wehling_PRB07,WehlingNL08,Wehling_CPL09,Wehling_Midgap_PRB09} but not
necessarily \cite{Wehling_PRB07,robinson08}. For impurity resonances
some $100$\,meV off the neutrality point the
conductivity should display a pronounced electron-hole
asymmetry \cite{robinson08} which is not observed in experiments.
So, it is not clear whether resonant impurity scattering
can be the main limiting factor in a
general case.

In this Letter, we build a first-principles theory of electron
scattering by {\it realistic} resonant impurities, such as various
organic molecules which are always present in exfoliated graphene
samples \cite{meyer2007,gass2008}. Combining the Boltzmann
equation approach and a
numerically exact Kubo formula consideration with first-principles parameters, we show that this class of impurities can limit electron
transport in typical exfoliated graphene samples and explain the
experimentally observed concentration dependence of the
conductivity.

Exfoliated graphene samples are contaminated with long polymer
chains \cite{meyer2007,gass2008}. Most important about these contaminants is their possibility 
to form a chemical bond to carbon atoms from the graphene sheet. To model such a situation we
carry out density functional theory (DFT)
calculations of graphene with adsorbed CH$_3$, C$_2$H$_5$,
CH$_2$OH (as simplest examples of different organic groups), as well as H and OH groups. From the resulting supercell
band structures we derive effective interaction parameters
entering a TB model and find that the exact chemical
composition is not essential: the parameters are very similar for all adsorbates except for the case of hydroxyl. This facilitates
us to obtain the effect of the contamination on the electron
conductivity.

An atomistic description of the graphene adsorbate systems is
achieved by DFT calculations within the generalized
gradient approximation (GGA) \cite{Perdew:PW91} on $3\times
3$-$9\times 9$ graphene supercells containing one impurity. Using the
Vienna Ab Initio Simulation Package (VASP) \cite{Kresse:PP_VASP}
with the projector augmented wave (PAW)
\cite{Bloechl:PAW1994,Kresse:PAW_VASP} basis sets, we
obtain fully relaxed adsorption geometries and
calculate the supercell band structures.

\begin{figure*}[tb]
\begin{minipage}{0.6\linewidth}
\centering
\includegraphics[width=.9\linewidth]{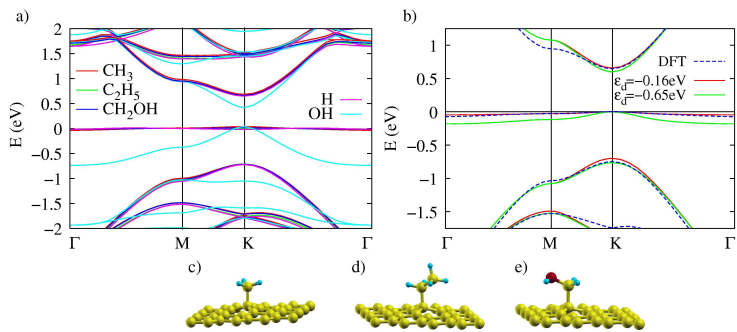}
\end{minipage}\hfill\begin{minipage}{0.39\linewidth}
\caption{\label{fig:bands}(Color online) (a) Band structures of
$4\times 4$ graphene supercells with CH$_3$, C$_2$H$_5$, CH$_2$OH,
H and OH adsorbates and the respective adsorption geometries of the
CH$_3$, C$_2$H$_5$, CH$_2$OH (c-e) groups.
(b) Comparison of the supercell band structure of graphene with CH$_3$
as obtained from DFT to the TB models with $V=2t$ and on-site
energies $\epsilon_d=-0.16$\,eV and $\epsilon_d=-0.65$\,eV. }
\end{minipage}
\end{figure*}

The DFT results for CH$_3$, C$_2$H$_5$, CH$_2$OH on graphene  are shown in Fig.
\ref{fig:bands}a and compared to H and OH adsorbates. All of these
impurities bind covalently to graphene and create a midgap state
as characteristic for monovalent impurities
\cite{Wehling_Midgap_PRB09}. For all adsorbates
except OH the midgap state lies within $\pm
0.03$\,eV around the neutrality point. As the supercell band
structures for the organic groups and for H on graphene
virtually coincide within an energy range of more than $\pm
1$\,eV, it becomes clear that the parameters of the midgap state
depend very weakly on the adsorbed group and, thus, can be
considered as robust for further use in the transport theory.

For an analytical description of these systems we start with a
TB model of graphene,
\begin{equation}
\hat{H}=-t\sum_{<i,j>}c^\dagger_i c_j, \label{ham}
\end{equation}
where $c_i$ denotes the Fermi operator of an electron in the
carbon $p_z$ orbital at site $i$, the sum includes all pairs of
nearest-neighbor carbon atoms, and $t\approx 2.6$ eV is the
nearest-neighbor hopping parameter. In this framework, we consider
a ``non-interacting Anderson impurity'', adding to the (\ref{ham})
 the localized state, $
\hat{H}_{\imp}=\epsilon_{d}d^\dagger d, $ with on-site energy
$\epsilon_{d}$ and corresponding Fermi operator $d$, which is
coupled to the graphene bands by $\hat{V}=V c^\dagger_0 d +
\rm{H.c.}$.

To describe electron transport in pristine as well as doped
graphene correctly, the analytical model has to recover the
realistic system within an energy window of some $100$\,meV around
the neutrality point. Applying the same supercell boundary
conditions as in the DFT simulations to the TB impurity model, we
obtain the TB supercell band structures as depicted in Fig.
\ref{fig:bands} b. The band structure of graphene with a methyl
group is well fitted with $V\approx 2t=5.2$\,eV and
$\epsilon_d\approx -t/16=-0.16$\,eV.

For the DFT band structures of all other neutral functional groups we
find a good fit of TB with $|V|\gtrsim 2t$ and
$|\epsilon_d|\lesssim 0.1t\approx 0.26$\,eV. The hybridization
strength $V$ being a factor 2 larger than $t$ is in accordance
with the hybridization for hydrogen ad-atoms from Ref.
\cite{robinson08} and appears very reasonable, as the impurity
forms a $\sigma$-bond with the host atom underneath \footnote{The
hybridization parameter $V$ should not be confused
with the avoided crossing strength from Ref.
\onlinecite{Wehling_Midgap_PRB09}. The latter is supercell
specific, in contrast to $V$ used here.}. The on-site
energies $|\epsilon_d|$ obtained here are significantly
smaller than the value $\epsilon_d=1.7$\,eV used for H in Ref. \cite{robinson08} which will make
our results for the transport properties qualitatively different.
We note that the model parameters extracted here are converged
w.r.t. the supercell size.

The scattering of electrons caused by resonant impurities is
described by the $T$-matrix (for a review, see Ref.
\onlinecite{Wehling_CPL09})
 $T(E)=\frac{V^2}{E-\epsilon_{d}-V^2g_0(E)}$,
where
$g_0(E)\approx\frac{E}{D^2}\ln\left|\frac{E^2}{D^2-E^2}\right|-i\pi
N_0(E)$, with $N_0(E)=\frac{|E|}{D^2}\cdot\Theta(D-|E|)$ and
$D=\sqrt{\sqrt{3}\pi}t\approx 6$\,eV, is the local Green function
of pristine graphene. Correspondingly, $N_0(E)$ is the density of
states (DOS) per spin and per carbon atom.
The $T$-matrix exhibits a resonance at
$E\left(1-\frac{V^2}{D^2}\ln\left|\frac{E^2}{D^2-E^2}\right|\right)-\epsilon_{d}=0$
which is the energy of the midgap state. The impurity model
parameters obtained from DFT lead to resonances in an energy
region of $\pm 0.03$\,eV around the Dirac point, which proves
consistency of our TB model with DFT.

In the Boltzmann equation approach,
the $T$-matrix can be used to estimate the conductivity $\sigma$: $\sigma=(2e^2/h)v_Fk_F\tau$, where $v_F$ is the Fermi
velocity and $k_F$ is the Fermi wave vector. For a concentration
of $n_i$ impurities per carbon atom, the scattering rate reads as
\cite{MahanBook,Ando_98,robinson08} $\tau^{-1}=(2\pi/\hbar)n_i
|T(E_F)|^2 N_0(E_F)$ and yields the conductivity
\begin{eqnarray}
\sigma&\approx&(2e^2/h)(2\pi n_i |T(E_F)/D|^2)^{-1}. \label{eqn:sigma_Boltzmann}
\end{eqnarray}
In the limit of resonant impurities with $V\to\infty$, we obtain $T\rightarrow -1/g_0(E)\approx
-\left[\frac{2E}{D^2}\ln\left|\frac{E}{D}\right|\right]^{-1}$ for
$E\ll D$. Hence, the conductivity reads in this limit as
\begin{eqnarray}
\sigma&\approx&(2e^2/h)\frac{2}{\pi}\frac{n_e}{n_i} \ln^2\left|\frac{E_F}{D}\right|, \label{eqn:imp_lim}
\end{eqnarray}
where $n_e=E_F^2/D^2$ is the number of charge carriers per carbon
atom. Eq. (\ref{eqn:imp_lim}) yields the same behavior as for
vacancies \cite{SPG07}. In the case of the resonance
shifted with respect to the neutrality point the consideration of
Ref. \cite{KN07} leads to the dependence
\begin{equation}
 \sigma \propto \left( q_0 \pm k_F\ln{k_F R} \right)^2, \label{fitting}
\end{equation}
where $\pm$ corresponds to electron and hole doping, respectively,
and $R$ is the effective impurity radius.

\begin{figure}[ht]
 \centering
\begin{minipage}{0.7\linewidth} \raggedright
  \includegraphics[width=.98\linewidth]{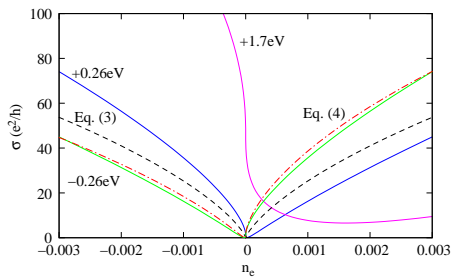}
\end{minipage}
\caption{\label{fig:sigma}(Color online) Conductivity $\sigma$ in the
Boltzmann approach as function of charge carrier concentration
$n_e$ (in units of electrons per atom) for different impurities:
Impurities with hybridization $V=2t=5.2$\,eV and on-site energies
$\epsilon_d=-0.26$, $0.26$, and $1.7$\,eV in concentration
$n_i=0.1\%$. (Curves are labelled by the corresponding
$\epsilon_d$.) Fits to the $V\to\infty$ limit of Eq.
(\ref{eqn:imp_lim}) with $n_i=0.06\%$ (dashed) as well as
Eq. (\ref{fitting}) with $q_0=0.02\,\ang^{-1}$ (dash dotted)
are shown. (Here, $n_e=E_F^2/D^2$ corresponds to the clean graphene DOS.)}
\end{figure}

We now investigate to which extend realistic
resonant impurities create sublinear behavior similar to Eqs.
(\ref{eqn:imp_lim}-\ref{fitting}). To this end, we first estimate
the conductivity according to Eq. (\ref{eqn:sigma_Boltzmann}) for
different types of impurities (Fig. \ref{fig:sigma}).
For the resonant scatterers from Fig. \ref{fig:bands}
(except for OH) the conductivity curves are expected to lie
within the region bounded by the curves belonging to
$\epsilon_d=-0.26$\,eV and $\epsilon_d=0.26$\,eV. These
curves are very similar to V-shape experimental
curves \cite{kostya2,kim,Mohiuddin09,Nietal10} and can be roughly
fitted to the limit of Eqs. (\ref{eqn:imp_lim}) and
(\ref{fitting}). The effective radius $R$ resulting from Eq.
(\ref{eqn:imp_lim}) is $R=D/{\hbar v_F}\approx 0.9\ang$ and has
been also used in the fit according to Eq. (\ref{fitting}) in Fig.
\ref{fig:sigma}. Experimentally, sublinear behavior similar to
Eqs. (\ref{eqn:imp_lim}-\ref{fitting}) has been observed
\cite{Chetal09,Nietal10} with effective impurity radii in the
range of $R=2.3-2.9\ang$. However, any estimation of effective
radii should be considered only qualitatively, as $D$ and
$R$ enter the conductivity logarithmically and a wide range of
cut-offs lead to similar conductivity curves.

The result for impurities with $V=2t$ and $\epsilon_d=1.7$\,eV,
which corresponds to H ad-atoms in the model of Ref.
\cite{robinson08}, differs qualitatively from our results and from
experimental data which emphasizes the crucial importance of a
careful {\it first-principles} determination of the model
parameters. In our model and for the charge carrier concentration
being varied within $|n_e|<0.003$/C-atom$=1.1\cdot
10^{13}$\,cm$^{-2}$, impurities like CH$_3$, C$_2$H$_5$,
CH$_2$OH, or H attached to graphene lead to a Boltzmann
conductivity with one distinct minimum close to the
neutrality point.

\begin{figure}[t]
\begin{center}
\includegraphics[width=0.9\linewidth]{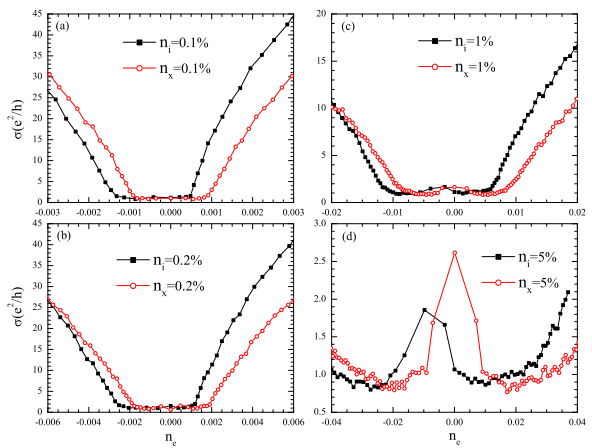}
\end{center}
\caption{\label{fig:cond_Kubo}(Color online) Conductivity $%
\protect\sigma $ as a function of charge carrier concentration $n_e$ (in
units of electrons per atom) for different resonant impurity ($\protect%
\varepsilon _{d}=-t/16,$ $V=2t$) or vacancy concentrations ($n_x$) :
(a) $n_i=n_x=0.1\%,$ (b) $0.2\%,$ (c) $1\%,$ (d) $5\%$.
Periodic boundary conditions are used with a sample containing (a) $8192\times 8192$ and (b-d) $4096\times 4096$
carbon atoms. The carrier concentrations $n_e$ are obtained from the integral of the corresponding DOS
depicted in Fig. (\ref{fig:dos}).}
\end{figure}

\begin{figure}[t]
\begin{center}
\includegraphics[width=0.9\linewidth]{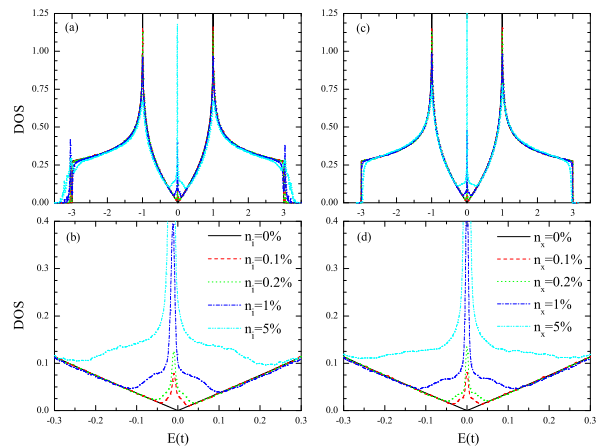}
\end{center}
\caption{\label{fig:dos}(Color online) Density of states as a function of energy $E$ for different resonant impurity ($\protect%
\varepsilon _{d}=-t/16,$ $V=2t$) or vacancy concentrations: $n_i (n_x) =0.1\%,$ $0.2\%,$ $1\%,$ $5\%$.}
\end{figure}

At low charge carrier concentrations or high impurity
concentrations, the Boltzmann approach becomes questionable. To
understand the on-set of this parameter regime and the behavior of
the conductivity in this regime, we performed numerically exact
calculations of the conductivity in the TB model (\ref{ham})
 using the Kubo formula. [See \cite{kubo}.]
The results for two types of resonant scatterers, adsorbed atoms
with $\epsilon_d=-t/16$, $V=2t$ resembling CH$_3$ groups, and
for vacancies are shown in Fig. \ref{fig:cond_Kubo}.
One can see that the Boltzmann equation is
applicable only for impurity concentrations smaller than a few
percent per site (already for 5\% the difference in concentration
dependence is essential). The Boltzmann approach does not work
near the neutrality point where quantum corrections are dominant
\cite{OGM06,Katsnelson_zitter,AK07}. In the range of
concentrations, where the Boltzmann approach is applicable the
conductivity as a function of energy fits very well the dependence
of Eq. (\ref{fitting}), with $q_0=0.02\ang^{-1}$, $R=0.6\ang$ for
$n_i=0.1\%$, and $q_0=0$, $R=0.5\ang$ for $n_x=0.1\%$ with $k_F=E_F/(\hbar v_F)$ as in clean graphene.

Close to the neutrality point the conductivity deviates from the
Boltzmann equation result of Eq. (\ref{eqn:sigma_Boltzmann}). Boltzmann theory is not capable of yielding
$\sigma=4e^2/\pi h$ for clean graphene at the neutrality point
\cite{OGM06,Katsnelson_zitter}. Moreover, 
resonant impurities lead to the formation of a low energy impurity
band (see increased DOS at low energies in Fig. \ref{fig:dos}). At impurity concentrations on the order of a few percent (Fig. \ref{fig:cond_Kubo} c,d) this impurity band contributes to the conductivity and can lead to a maximum of $\sigma$ in the midgap region. Moreover, the impurity band can host two electrons per impurity. For impurity concentrations below $\sim 5\%$, this leads to a plateau shaped minimum of width $2n_i$ (or $2n_x$) in the conductivity vs. $n_e$ curves around the neutrality point. Analyzing the plateau width in experimental data (similar to the analysis for N$_2$O$_4$ acceptor states in Ref. \onlinecite{WehlingNL08}) can, thus, yield an independent estimate of impurity concentrations. For chiral disorder \cite{Altland02,Mirlin10} corresponding to the resonant impurities considered, here, as well as short range disorder \cite{Roche08a,Roche08b} (anti)localization effects can become important in cases like graphane, where impurity concentrations are varied between a few percent and $100\%$. In clean micron size samples with realistic impurity concentrations on the order of $n_i=0.01\%-0.1\%$ these effects present merely corrections: Upon doubling the simulation cell length ($4096\times4096\,\to\,8192\times 8192$) at $n_i=0.1\%$ the changes of the conductivity at the neutrality point are below $10\%$.

Electron scattering in bilayer graphene has been proven to differ essentially from the single layer case in Ref. \onlinecite{Katsnelson_bilayer}:
For a scattering potential with radius much smaller than the de Broglie
wavelength of electrons, the phase shift of $s$-wave scattering $\delta_0$ tends to a constant as $k\to 0$. Therefore, within the limit
of applicability of the Boltzmann equation, the conductivity of a bilayer should be just linear
in $n_e$, instead of sublinear dependence (\ref{fitting}) for the single layer. The difference is that in the single layer,
due to vanishing DOS at the Dirac point, the scattering disappears at small wave vectors as
$\delta_0(k) \propto \frac{1}{\ln{kR}}$ (with $\ln^2{kR}$ on the order of 10 for typical amounts of doping) for resonant and as $\delta_0(k) \propto kR$ for the nonresonant
impurities. Contrary, in the bilayer there are no restrictions on the strength of the scattering and even the
unitary limit ($\delta_0 = \pi/2$) can be reached at $k=0$. As follows from Ref. \onlinecite{Katsnelson_bilayer}, a cylindric potential well of radius $R$, leads to $\delta_0 = \pi/2$ if $\frac{d}{dR} \frac{J_0 (qR)}{I_0 (qR)} =0$,
where $q$ is the wave vector inside the well, $J_0$ and $I_0$ are the Bessel functions of real and
imaginary arguments, respectively. Thus, an assumption that resonant scattering is the main limiting factor for electron mobility in
exfoliated graphene leads to the prediction that the dependence of $\sigma(n_e)$ should be
essentially different for the cases of bilayer and single layer, that is, linear and sublinear, respectively. This agrees with the experimental results \cite{morozov08}.

In summary, we have demonstrated that realistic impurities in graphene 
frequently cause quasilocal peaks nearby the neutrality point. In
particular, for various organic groups the formation of a
carbon-carbon bond results in the appearance of midgap
(resonant) states within $\pm 0.03$\,eV around the neutrality point. They can be
described as Anderson impurities with the hybridization parameter
of about 2$t$ and on-site energies on the order of $|\epsilon_d|<t/10$.
The resonant scattering model with these parameters
describes satisfactory experimental data on the concentration
dependence of charge carrier mobility for graphene.

We thank L. Oroszl{\'a}ny and H. Schomerus for discussions of Ref. \onlinecite{robinson08}. Support from
SFB 668 (Germany), the Cluster of Excellence "Nanospintronics" (LExI Hamburg), FOM (The Netherlands) and computer time
at HLRN (Germany) and NCF (The Netherlands) are acknowledged.
\bibliography{water_s}

\end{document}